\documentclass[aps,prd,twocolumn,superscriptaddress,nofootinbib]{revtex4}

\usepackage{graphicx}
\usepackage{dcolumn}
\usepackage{bm}
\usepackage{amsmath,amsfonts,amssymb}
\usepackage{epsfig}    
\usepackage{color}
\usepackage{slashed}
\usepackage{hhline}
\usepackage{newtxtext,newtxmath}
\usepackage{mathrsfs}
\usepackage{longtable}
\usepackage{natbib}
\usepackage{setspace}
\usepackage{xspace}
\usepackage[utf8]{inputenc}

\def\be{\begin{equation}}
\def\ee{\end{equation}}
\newcommand{\bea}{\begin{eqnarray}}
\newcommand{\eea}{\end{eqnarray}}
\newcommand{\nn}{\nonumber}

\begin{document}

\title{
  \begin{flushright}
    \rightline{KIAS-P19002, APCTP Pre2019-001}
  \end{flushright}
  Dark matter and $B$-meson anomalies in a flavor dependent gauge symmetry}

\author{Parada~T.~P.~Hutauruk}
\email{parada.hutauruk@apctp.org}
\affiliation{Asia Pacific Center for Theoretical Physics, Pohang, Gyeongbuk 37673, Korea }

\author{Takaaki Nomura}
\email{nomura@kias.re.kr}
\affiliation{School of Physics, KIAS, Seoul 02455, Korea}

\author{Hiroshi Okada}
\email{hiroshi.okada@apctp.org}
\affiliation{Asia Pacific Center for Theoretical Physics, Pohang, Gyeongbuk 37673, Korea }

\author{Yuta Orikasa}
\email{Yuta.Orikasa@utef.cvut.cz}
\affiliation{Institute of Experimental and Applied Physics, Czech Technical University,
  Prague 12800, Czech Republic}

\date{\today}

\begin{abstract}
  A possibility of explaining the anomalies in the semileptonic $B$-meson decay
  $B \to K^{*} \mu \bar\mu$ has been explored in the framework of
  the gauged $U(1)_{\mu-\tau}$ symmetry.
  Apart from the muon anomalous magnetic moment and neutrino sector,
  we formulate the model starting with a valid Lagrangian
  and consider the constraints from the neutral meson mixings,
  the bounds on direct detection and the relic density of the bosonic
  dark matter candidate augmented to collider constraints.
  We search the parameter space, which accommodate the size of the anomaly of the
  $B \rightarrow K^* \mu \bar \mu$ decay, to satisfy all experimental constraints.
  We found the allowed region on the plane of the dark matter and $Z'$ masses is rather narrow
  compared to the previous analysis.
\end{abstract}
\maketitle

\section{Introduction} \label{sec:intro}
A flavor dependent gauge symmetry is one of the promising candidates for new physics
to describe the anomalies and other phenomenologies related with the flavor physics
as well as to ensure the dark matter (DM) stability.
In particular, the model of $U(1)_{\mu-\tau}$ provides several phenomenological
prescriptions to resolve, namely, muon anomalous magnetic moment~\cite{Altmannshofer:2014pba},
\footnote{Recently, a stringent constraint of the neutrino-trident
  process gives narrower parameter spaces of the extra gauge coupling ($g'$)
  and mass ($m_{Z'}$).} experimental anomalies of semileptonic
$B$-meson decay~\cite{Ko:2017yrd,Arcadi:2018tly},
neutrino sector~\cite{Nomura:2018vfz,Nomura:2018cle,Lee:2017ekw,Baek:2015mna,
  Dev:2017fdz,Baek:2017sew,Chen:2017gvf,Asai:2017ryy,Chen:2017cic,Biswas:2016yan,
  Baek:2015fea,Asai:2018ocx}, and other related topics~\cite{Banerjee:2018mnw,Heeck:2018nzc}.
Among them, the $B$-meson decay anomaly is a very challenging topic
due to some indications of new physics have been suggested in the $B$ physics.
For example, the angular observable $P'_5$ in the decay of the $B$ meson 
$B \to K^* \mu^+ \mu^-$~\cite{DescotesGenon:2012zf} has been measured
with deviation of $3.4 \sigma$ from the integrated luminosity of 3.0 fb$^{-1}$
at the LHCb~\cite{Aaij:2015oid} which confirms the previous 
result with deviation of $3.7\sigma$~\cite{Aaij:2013qta}.
In addition, the same observable were measured by Belle collaboration~\cite{Abdesselam:2016llu,Wehle:2016yoi}
with the deviation of $2.1 \sigma$. 
Furthermore, an anomaly in the measurement of the ratio of
branching fraction $R_K = BR(B^+ \to K^+ \mu^+\mu^-)/ BR(B^+ \to K^+ e^+e^-)$~\cite{Hiller:2003js,
  Bobeth:2007dw} at the LHCb indicates a deviation of $2.6\sigma$ from the lepton universality
predicted in the standard model (SM)~\cite{Aaij:2014ora}.  
Recently, the LHCb collaboration has also measured the ratio
of $R_{K^*} = BR (B \to K^* \mu^+ \mu^-)/ BR (B \to K^* e^+ e^-)$
which is found to be deviated from the SM prediction by $\sim 2.4 \sigma$
as $R_{K^*} = 0.660^{+ 0.110}_{- 0.070} \pm 0.024 (0.685^{+ 0.113}_{- 0.069} \pm 0.047)$
for $(2 m_\mu^2) < q^2 < 1.1$ GeV$^2$ (1.1 GeV$^2 < q^2 < 6$ GeV$^2$)~\cite{Aaij:2017vbb}.

In previous study of Ref.~\cite{Ko:2017yrd}, we have proposed the flavor dependent gauge symmetry.
This has successfully explained the anomaly of $B \to K^{*} \ell^{+} \ell^{-}$ decay
through generating the flavor violating $Z'$ boson interactions at one loop level.
In the model, the only Wilson coefficient $C_9$ with $\mu$ and $\tau$ in the $B$ decays
are generated via extra $Z'$ boson exchange, but this is explicitly not applicable for
$B \to K^{*} e \bar e$ process~\cite{DescotesGenon:2012zf}.
We have also included a bosonic DM $\chi$ candidate and the vectorlike exotic quarks $Q'_a$,
which are needed to generate the Wilson coefficient $C_9$ at one loop level.
Thus the DM relic density~\cite{Ade:2013zuv} can be explained the measured anomalies
in decay of $B \to K^{*} \ell^{+} \ell^{-}$ via s-channel process mediated by $Z'$ boson
exchange~\cite{Sierra:2015fma, Belanger:2015nma},
where $Z'$ boson exchange can avoid a conflict with the constraints from
the spin independent DM direct detection searches
such as experiments of LUX~\cite{Akerib:2016vxi} and XENON1t~\cite{Aprile:2018dbl}.

In this paper, we adopt the flavor dependent gauge symmetry model with more complete manner.
We then perform a more detailed analysis in which
we take into account the decay width of the $Z'$ boson
which is related with the relic density of the DM, the constraints
from the spin independent DM-nucleon elastic scattering cross section
mediated by the vectorlike quarks $Q'_a$ at tree level,
the large electron-positron (LEP) collider, and the large hadron collider (LHC),
to improve the previous analyses of Ref.~\cite{Ko:2017yrd}.
In our present numerical analysis, we find that the allowed regions of the DM
and $Z'$ masses are narrower than that of the previous analysis.
This is expected due to the decay width of $Z'$ is rather larger.

This letter is organized as follows.
In Sec.~\ref{sec:modelsetup}, we briefly introduce a valid Lagrangian of
our model including the Higgs potential with the inert conditions,
the $B$-meson anomaly, the collider physics, and the neutral meson mixings.
A brief comment on how to directly produce the exotic quarks,
and the experimental constraints of the DM are also presented.
In Sec.~\ref{sec:numerical}, we present our numerical analysis results.
Finally Sec.~\ref{sec:summary} is devoted to the summary of our results and conclusions.
%
%

\section{Model setup and constraints} \label{sec:modelsetup}
%
\begin{center} 
\begin{table}[b]
\caption{Field contents of fermions and their charge assignments under
$SU(2)_L \times U(1)_Y \times  U(1)_{\mu-\tau}$, where $q_x \neq 0$ and
the lower index $a$ of $Q'$ is the number of family which runs over $1-3$.}
\label{tab:1}
\begin{footnotesize}
\begin{tabular}{|c||c|c|c|c|c|c|c|c|}\hline\hline  
& \multicolumn{6}{c|}{Leptons} & \multicolumn{1}{p{15mm}|}{Exotic vector fermions} \\\hline
Fermions  ~&~ $L_{L_e}$ ~&~ $L_{L_\mu}$ ~&~ $L_{L_\tau}$ ~&~ $e_R$ ~&~ $\mu_R$ ~&~ $\tau_R$ ~&~ $Q'_{a}$~
\\\hline 
$SU(3)_C$  & $\bm{1}$  & $\bm{1}$  & $\bm{1}$   & $\bm{1}$  & $\bm{1} $  & $\bm{1}$ & $\bm{3}$  \\\hline 
$SU(2)_L$  & $\bm{2}$  & $\bm{2}$  & $\bm{2}$   & $\bm{1}$  & $\bm{1}$   & $\bm{1}$  & $\bm{2}$  \\\hline 
$U(1)_Y$  & $-\frac{1}{2}$ & $-\frac12$ & $-\frac12$  & $-1$ &  $-1$  &  $-1$  & $\frac16$  \\\hline
$U(1)_{\mu-\tau}$ & $0$  & $1$ & $-1$ & $0$  & $1$   & $-1$ & $q_x$  \\\hline
\end{tabular}
\end{footnotesize}
\end{table}
\end{center}
%

\begin{table}[t]
\caption{Field contents of bosons and their charge assignments under
$SU(2)_L \times U(1)_Y \times U(1)_{\mu-\tau}$, where $q_{x,y} \neq 0$, $SU(3)_C$
singlet for all bosons, and $\chi$ is a complex boson which is considered as a DM candidate.
Hence $q_y \neq \pm q_x$, $q_y \neq \pm 2 q_x$, and $q_x \neq \pm 3 q_y$
are simultaneously satisfied.}
\label{tab:2}
\addtolength{\tabcolsep}{10.8pt}
\centering {\fontsize{10}{12}
\begin{tabular}{|c||c|c||c|c|}\hline\hline
&\multicolumn{2}{c||}{VEV$\neq 0$} & \multicolumn{1}{c|}{Inert } \\\hline
Bosons  &~ $H$  ~ &~ $\varphi$ ~ &~ $\chi$   ~ \\\hline
$SU(2)_L$ & $\bm{2}$ & $\bm{1}$ & $\bm{1}$    \\\hline 
$U(1)_Y$ & $\frac12$ & $0$ & $1$     \\\hline
$U(1)_{\mu-\tau}$ & $0$  & $q_y$ & $q_x$  \\\hline
\end{tabular}%
} 
\end{table}

{In this section, we present a formulation of our model.
We briefly introduce a gauged $U(1)_{\mu-\tau}$ symmetry with
three families of the vectorlike isospin doublet quarks $Q'_a$,
an isospin singlet inert complex boson $\chi$,
and singlet boson $\varphi$ with nonzero vacuum expectation value (VEV)
which is denoted by $\langle \varphi \rangle \equiv v_\varphi/ \sqrt2$,
where $H$ is the SM Higgs and its VEV is denoted by
$\langle H \rangle \equiv v_H/ \sqrt2$.
The charge assignments of these new fermion and boson fields
are summarized in Tables~\ref{tab:1} and~\ref{tab:2}, respectively.

A relevant Lagrangian under these symmetries is defined by 
\begin{align}
-\mathscr{L} = y_{\ell_i} \bar L_{L_i} H e_{R_i}
+  f_{aj} \overline{Q'_{R_a}} Q_{L_j} \chi 
+&  M_a \bar Q'_a Q'_a + {\rm h.c.},
\label{Eq:lag-flavor}
\end{align}
where $a=1-3$ and $(i,j)= e, \mu, \tau$ are generation indices, and 
the quark sector $Q_{L_{j}}$ is same as the SM. Note that the charged-lepton sector is
diagonal due to the $U(1)_{\mu-\tau}$ symmetry.

\noindent \underline{\it Higgs potential} is given by
\begin{align}
V &= m_H^2 |H|^2 +m_\varphi^2 |\varphi|^2
+ m_\chi^2 |\chi|^2 + \frac14\lambda_{H} |H|^4
+ \frac14\lambda_\varphi |\varphi|^4\nn\\
&+ \frac14 \lambda_\chi |\chi|^4
+ \lambda_{H\varphi} |H|^2|\varphi|^2
+ \lambda_{H\chi} |H|^2|\chi|^2
+ \lambda_{\varphi\chi} |\varphi|^2|\chi|^2,
\label{Eq:pot}
\end{align}
where each field is defined to be 
\begin{align}
& H = \begin{pmatrix} h^+ \\
\frac{1}{\sqrt{2}} (v_H + h + i z) \end{pmatrix}, \quad
\varphi = \frac{1}{\sqrt{2}} (v_\varphi + \varphi_R + i z'),\quad
\end{align}
where $h^+$, $z$, and $z'$ are respectively absorbed by
the gauged bosons; $W^+$, $Z$, and $Z'$.
After inserting the tadpole conditions for $H$ and $\varphi$,
the CP-even mass matrix is obtained as
\begin{align}
M^2_{even} = \begin{pmatrix} v_H^2 \lambda_H
&  2 v_H v_\varphi \lambda_{H\varphi}  \\
2 v_H v_\varphi \lambda_{H\varphi}
& v_\varphi^2 \lambda_{\varphi} \end{pmatrix}. \label{eq:mat}
\end{align}
After the diagonalization,
the mass eigenvalues and eigenstates are respectively given by
\begin{align}
& O M^2_{even} O^T=
\begin{pmatrix} c_\theta & -s_{\theta}  \\
s_{\theta} & c_{\theta} \end{pmatrix}
\begin{pmatrix} v_H^2 \lambda_H
&  2 v_H v_\varphi \lambda_{H\varphi}  \\
2 v_H v_\varphi \lambda_{H\varphi}
& v_\varphi^2 \lambda_{\varphi} \end{pmatrix}
\begin{pmatrix} c_\theta & s_{\theta}  \\
- s_{\theta} & c_{\theta} \end{pmatrix}\nn\\
&={\rm diag}(m^2_{h_{SM}}, m^2_{H}),\\
& m^2_{h_{SM}} = \frac12 \left(A -\sqrt{(v_\varphi^2 \lambda_{\varphi}- v_H^2 \lambda_{H})^2  +16 v_\varphi^2 v_H^2 \lambda_{H\varphi}^2 } \right),\nn\\
& m^2_{H} = \frac12 \left(A + \sqrt{(v_\varphi^2 \lambda_{\varphi}- v_H^2 \lambda_{H})^2 +16 v_\varphi^2 v_H^2 \lambda_{H\varphi}^2 } \right),\nn
\end{align}
where $A =v_\varphi^2 \lambda_{\varphi}+ v_H^2 \lambda_{H}$, $\sin\theta(\cos\theta)\equiv c_\theta(s_\theta)$, and $\tan \theta \equiv t_\theta$ satisfies the following relation:
\begin{align}
2 v_H v_\varphi \lambda_{H\varphi} t_\theta^2 -(v_\varphi^2\lambda_\varphi-v_H^2 \lambda_{H}) t_\theta -2v_H v_\varphi \lambda_{H\varphi} = 0.
\end{align}
The mass eigenvalue of $\chi$ is given by
\begin{align}
m^2_{\chi} = v_\varphi^2 \lambda_{\varphi\chi}
+ v_H^2 \lambda_{H\chi}
+ 2m_\chi^2. 
\end{align}
{The inert conditions for $\chi$ are given by 
\begin{align}
\lambda_H, \lambda_\varphi, \lambda_\chi, \lambda_{H\varphi}, \lambda_{H\chi}, \lambda_{\varphi\chi} &> 0, 
\nonumber\\
m_\chi^2 & > 0,  
\nonumber\\
\lambda_H \lambda_\varphi & > 4 \lambda_{H\varphi}^2. 
\end{align}
}

\noindent \underline{\it $Z'$ boson}:
We have $Z'$ boson from $U(1)_{\mu - \tau}$ gauge symmetry.
After $\varphi$ develops its VEV, the mass of $Z'$ boson is generated as 
\begin{equation}
m_{Z'} = q_y^2 g'^2 v_\varphi^2,
\end{equation}
where $g'$ is a gauge coupling for $U(1)_{\mu - \tau}$.
The gauge interactions among $Z'$ and fermions are given by
\begin{equation}
  \begin{split}
\mathscr{L} &\supset g' Z'_\mu ( \bar L_{L_\mu} \gamma^\mu L_{L_\mu}
- \bar L_{L_\tau} \gamma^\mu L_{L_\tau} 
+ \bar \mu_{R} \gamma^\mu \mu_R \\
&- \bar \tau_{R} \gamma^\mu \tau_R
+ q_y \sum_{a=1}^3 \bar Q'_a \gamma^\mu Q'_a ).
  \end{split}
\end{equation}
Note that we ignore the kinetic mixing effects between $U(1)_{\mu - \tau}$ and $U(1)_Y$
by assuming the contributions is relatively very small.

\noindent \underline{\it Explanation of the anomaly in $B \to K^{(*)} \ell^+ \ell^-$ decay}:
In our case, we have a Wilson coefficient $C_9$, which is associated with
$(\bar s\gamma_\mu P_L b)(\mu\gamma^\mu \mu)$, via Fig.~\ref{fig:diagram}.
Then their contribution to a Wilson coefficient $\Delta C_9^{\mu \mu}$
is given by~\cite{Ko:2017yrd, DescotesGenon:2012zf}
\begin{align}
\Delta C_9^{\mu \mu} & \simeq \frac{q_x g'^2}{(4\pi)^2 m_{Z'}^2 C_{\rm SM}}
\sum_{a=1}^{3} f^\dag_{3a} f_{a2}
\int[dx] \ln \left(\frac{\Delta[M_a,m_\chi]}{\Delta[m_\chi,M_a]}\right), \nn \\
C_{\rm SM} &\equiv \frac{V_{tb} V^*_{ts} G_{\rm F} \alpha_{\rm em}}{\sqrt2 \pi}, \label{eq:c9-formula} \\
\Delta[m_1,m_2] &= (x + y-1)(x m_b^2 + y m_s^2) 
+ x m_1^2 +( y + z ) m_{2}^2, \nn
\end{align}
where $[dx] \equiv \int_0^1 dx dy dz \delta(1-x-y-z)$, $V_{tb}$, $V_{ts}$ are
the 3-3 and 3-2 elements of the Cabibbo-Kobayashi-Maskawa (CKM) matrix, respectively,
$G_{\rm F} \approx 1.17 \times 10^{-5}~[\rm{GeV}]^{-2}$ is the Fermi constant,
$\alpha_{\rm em} \approx 1/137$
is the electromagnetic fine-structure constant, and
$m_b\approx4.18$ GeV and $m_s\approx0.095$ GeV are the bottom and strange quark masses,
respectively, which are given by the $\overline{\rm{MS}}$
renormalization scheme at a scale $\mu = 2$ GeV~\cite{Agashe:2014kda}.
We assume $m_b, m_s \ll m_{Z'}$ in Eq.~(\ref{eq:c9-formula}).
$m_{\chi}$ and $M_a$ are respectively $\chi$ and $Q'_a$ masses. 
We use the global fit for the value of $C_9$~\cite{Descotes-Genon:2015uva},
which gives the best fit value of $\Delta C_9 \approx $ -1.09.
The possible values of the $\Delta C_9$ are respectively obtained as 
\begin{equation}
\Delta C_9 = [- 1.29,-0.87]\ {\rm for}\ 1\sigma,\quad
[- 1.67,-0.39]\ {\rm for}\ 3\sigma \label{eq:C9_fit}.
\end{equation}

\begin{figure}[t]
\centering
\begin{tabular}{cc}
\begin{minipage}{0.5\hsize}
\includegraphics[width=40mm]{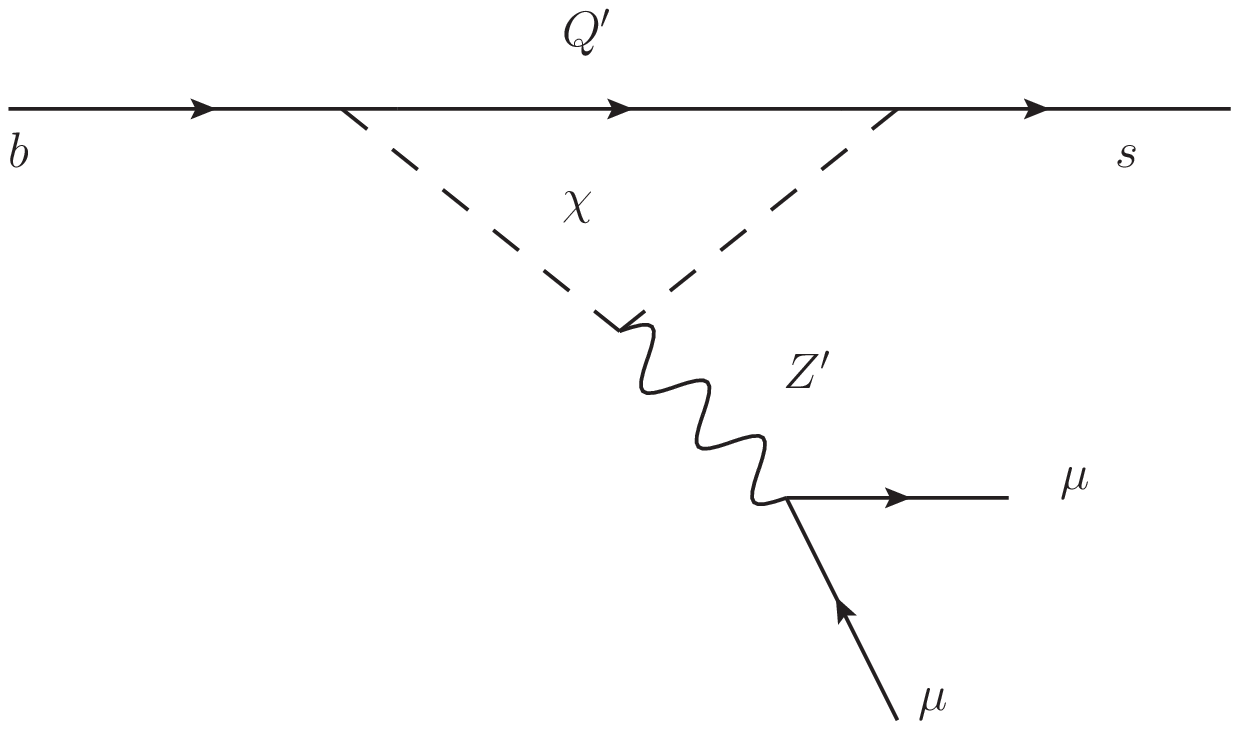} 
\end{minipage}
\begin{minipage}{0.5\hsize}
\includegraphics[width=40mm]{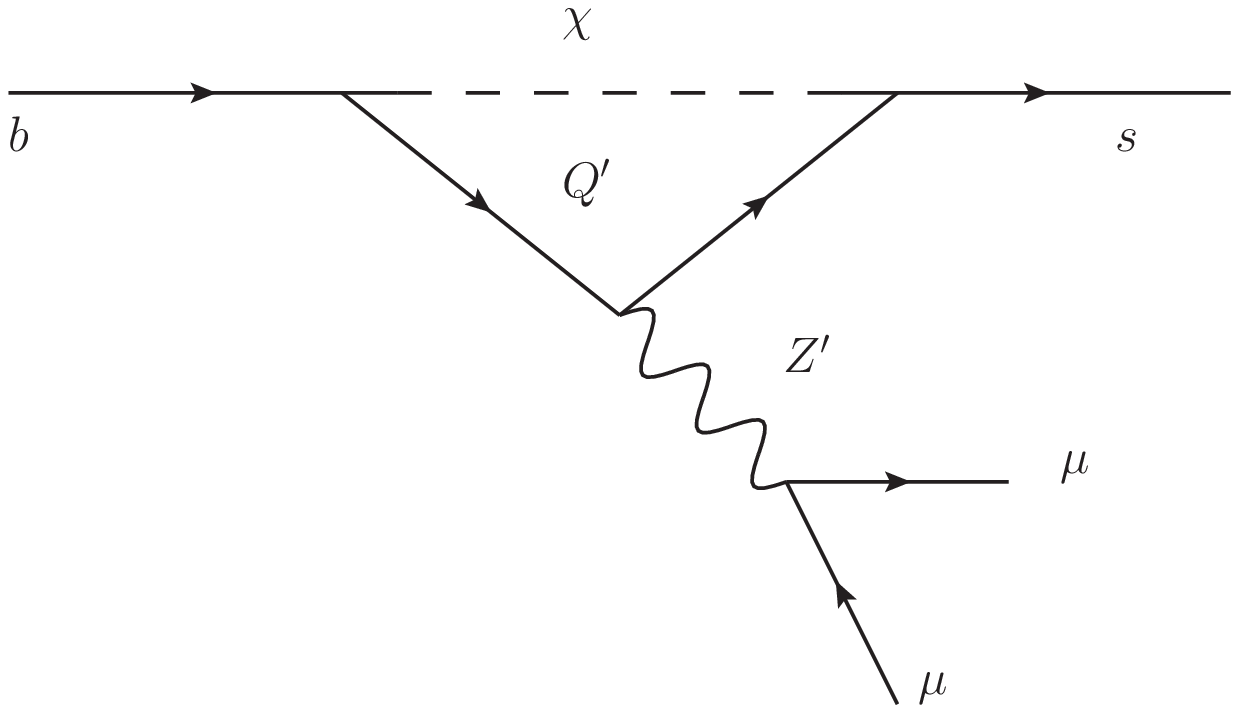} 
\end{minipage}
\end{tabular}
\caption{The diagrams introduces an effective coupling for
$Z'_\mu \bar b \gamma^\mu s + h.c.$ interaction.} 
\label{fig:diagram}
\end{figure}

\noindent
\underline{\it Constraint from $K$-meson decay}:
Here we discuss a constraint from $K$-meson decay.
Considering a small effect of the charge parity (CP) violation emerges from new physics,
the strongest bound is derived by $K_S \to \pi^0 \mu \bar \mu$ decay.
It is then induced by the effective Hamiltonian~\cite{Cirigliano:2011ny, Crivellin:2016vjc}
\begin{equation}
H_{eff} =  - \frac{G_F}{\sqrt{2}} V_{ud} V^*_{us} C_{7V}^{\mu \mu}
(\bar s \gamma^\mu (1- \gamma_5) d) (\bar \mu \gamma_\mu \mu).
\end{equation}
Similar to the Wilson coefficient $C_9$ case, the Wilson coefficient $\Delta C_{7V}$
arising from the $Z'$ exchange can be induced at one loop level. We then obtain
\begin{equation}
  \begin{split}
\Delta C_{7V}^{\mu \mu} &\simeq \frac{q_x g'^2}{(4\pi)^2 m_{Z'}^2} \frac{1}{\sqrt{2} G_F V_{ud} V_{us}^*} \\
&\times \sum_{a=1}^{3} f^\dag_{2a} f_{a1} \int[dx]\ln\left(\frac{\Delta[M_a,m_\chi]}{\Delta[m_\chi,M_a]}\right).
  \end{split}
\end{equation}
The values of the experimental constraint is given by~\cite{Cirigliano:2011ny} 
\begin{equation}
|a_S^{\mu \mu}| = 1.54^{+0.40}_{-0.32}, \quad a_S^{\mu \mu} \equiv \frac{2 \pi}{\alpha_{em}} V_{ud} V_{us}^* C_{7V}^{\mu \mu},
\end{equation}
where a large uncertainty is found. For estimating the $Z'$ contribution, we derive 
\begin{equation}
  \begin{split}
  \Delta a_S^{\mu \mu} & \simeq 0.46 \left( \frac{g'}{0.1} \right)^2 \left( \frac{100 \ {\rm GeV}}{m_{Z'}} \right)^2 \\
&\times  \sum_{a=1}^{3} f^\dag_{2a} f_{a1} \int[dx] \ln\left(\frac{\Delta[M_a,m_\chi]}{\Delta[m_\chi,M_a]}\right).
  \end{split}
\end{equation}
We find that $\Delta a_S^{\mu \mu}$ sufficiently {lies} within the range of the experimental uncertainty
if the Yukawa coupling $f$ is taken $O(0.1)$ and the constraint from $\Delta m_K$ tends to be stronger.

\noindent \underline{\it LHC constraint}: 
Since we consider no interactions between the electron-positron pair and $Z'$,
we evade the stringent constraint from the LEP experiment.
However, we need a constraint from the experiment of LHC,
while we consider interactions among u and d quarks via one-loop contributions,
as it can be explicitly seen in the part of $b \to s \ell \bar\ell$.

The most stringent constraint from the LHC arises from the process of
$q \bar q \to Z' \to \bar \mu \mu$. From this constraint,
the effective mass bound suggests 30 TeV $\lesssim \Lambda$, where its effective operator
is given by $\frac1{\Lambda^2}(\bar q \gamma_\mu P_L q)(\bar{\mu} \gamma^\mu P_L \mu)$~\cite{Aaboud:2017buh}.
We then can easily estimate the bound on $m_{Z'}$ in the effective operator analysis. 
Similar to the case of $b \to s \bar\mu \mu$, our effective operator is defined as
\begin{align}
\mathscr{L} &\simeq \frac{q_x g'^2}{(4\pi)^2 m_{Z'}^2} \sum_{a=1}^{3} f^\dag_{1a} f_{a1} \int[dx]
\ln \left(\frac{x M_a^2 + (y+z) m_{\chi}^2} {x m_\chi^2 + (y+z) M_{a}^2} \right) \nn \\
&\times (\bar q \gamma_\mu P_L q)(\bar{\mu} \gamma^\mu P_L \mu),
\label{eq:c9}
\end{align}
where $q = (u,d)$. It implies that we have the following constraint from the LHC:
\begin{align}
\frac{m_{Z'}}{g'} &\gtrsim 30 \left[ \frac{q_x}{(4 \pi)^3} \sum_{a=1}^{3} f^\dag_{1a} f_{a1} \int[dx]
  \ln\left(\frac{x M_a^2 +(y+z) m_{\chi}^2} {x m_\chi^2 +(y+z) M_{a}^2}\right)  \right]^{\frac12} \nn\\
     &\times {\rm TeV}.
\end{align}
When we take a degenerate mass for $M_a$ as $1000$ GeV and $m_\chi = 100$ GeV, the constraint then gives 
\begin{equation}
\frac{m_{Z'}}{g'} \gtrsim 0.45 \left[ \sum_{a=1}^3 f^\dag_{1a} f_{a1} \right]^{\frac12} \ {\rm TeV}.
\end{equation}
This shows that the constraint can be easily avoided while the coupling $f_{1a}$ is not too large.

\noindent \underline{\it $M-\overline M$ mixing}:  
%
%
\begin{figure}[t]
\begin{center}
\includegraphics[width=65mm]{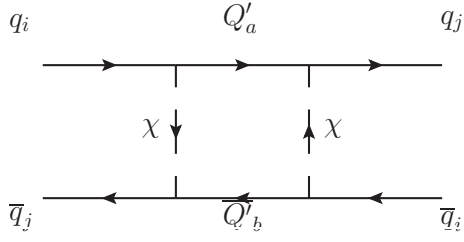} \qquad
\caption{The diagram representing the neutral meson mixings. } 
\label{fig:mesonmixing}
\end{center}\end{figure}
The neutral meson mixings also give the constraints of the parameter space. 
{The neutral meson mixings are shown in Fig.~\ref{fig:mesonmixing}} 
and their {formulas} be lower than the experimental bounds as follows
~\cite{Gabbiani:1996hi,Agashe:2014kda}:
\begin{align}
\Delta m_K &\approx
\sum_{a,b=1}^3
{f_{2a}^\dagger f_{a1} f_{2b}^\dagger f_{b1}} F^K_{box}[M_a,M_b] \nn \\  
& \lesssim 3.48\times10^{-15} \ [{\rm GeV}],
\label{eq:kk}\\
\Delta m_{B_d} &\approx
\sum_{a,b=1}^3
{f_{3a}^\dagger f_{a1} f_{3b}^\dagger f_{b1}}  F^{B_d}_{box}[M_a,M_b] \nn \\
& \lesssim 3.36\times10^{-13} \ [{\rm GeV}],\\
\Delta m_{B_s} &\approx
\sum_{a,b=1}^3
{f_{3a}^\dagger f_{a2} f_{3b}^\dagger f_{b2}}  F^{B_s}_{box}[M_a,M_b] \nn \\
& \lesssim 1.17\times10^{-11} \ [{\rm GeV}],\\
\Delta m_D &\approx
\sum_{a,b=1}^3
{f_{1a}^\dagger f_{a2} f_{1b}^\dagger f_{b2}}  F^{D}_{box}[M_a,M_b] \nn \\
&\lesssim 6.25\times10^{-15} \ [{\rm GeV}],\label{eq:dd} \\
& F^M_{box}(m_1,m_2)  =
\frac{m_M f_M^2}{3(4\pi)^2}
\int_0^1 \frac{x[dx]}{x m_\chi^2+y m_1^2+z m^2_2},\label{eq:mmbar-mix}
\end{align}
where $m_{M}$ and $f_M$ are the meson mass and the meson decay constant, respectively.
The following parameter values are used in our analysis: $f_K \approx 0.156$ GeV,
$f_{B_d(B_s)} \approx 0.191(0.274)$ GeV~\cite{DiLuzio:2017fdq, DiLuzio:2018wch},~\footnote{We
  thank Alexander Lenz to bring up a new value of $f_{B_s}$, which includes a bag parameter dependence.} $f_{D} \approx 0.212$ GeV,
$m_K \approx 0.498$ GeV, $m_{B_d(B_s)} \approx 5.280(5.367)$ GeV, and
$m_{D} \approx 1.865$ GeV.

\noindent \underline{\it Constraints from direct production of $Q'$s}:
The exotic quarks $Q'$s can be produced in pair via QCD processes at the LHC.
Each $Q'$ will then decay into $Q' \rightarrow q_i \chi$,
where $q_i$ represents a quark with flavor $i$.
Hence, searching for ``$\{ t t, b b , t  j, b j , j j  \}$ + missing $E_T$" signals
will constrain our present model.
The branching ratios for a particular quark flavor $i$ depend on
the relative sizes of the Yukawa couplings, $f_{3j}$ and $f_{aj}$ with $a = 1, 2$.  
We then roughly estimate the lower limit on the mass of $Q'$ using
the current LHC data for the s-quark searches~\cite{CMS:2016mwj, Aaboud:2016zdn},
which indicates the mass should be larger than $\sim 0.5$-$1$ TeV
which depends on the mass difference between $Q'$ and $\chi$.
In our following analysis, we simply take the value of $M_{a} > 1$ TeV
in order to satisfy this constraint.

\noindent \underline{\it Dark matter} : 
In our model scenario, a complex scalar $\chi$ is considered as
a DM candidate, since we have a remnant $Z_2$ symmetry after the $\mu-\tau$ symmetry breaking. 
The DM candidate and $Q'_a$ are odd under the $Z_2$ symmetry and the other particles are even. 
If the $M_a$ is heavier than the $m_\chi$, the DM candidate is a stable particle.
The DM dominantly annihilates into the SM leptons 
via $\chi \chi \to Z' \to \mu^+ \mu^- (\tau^+ \tau^-)$,~\footnote{
Notice that we do not rely on the Higgs portal, although there are two resonant solutions
at around the half masses of the SM Higgs and
another neutral Higgs~\cite{Kanemura:2010sh}.} so that the DM in our model 
is naturally leptophilic. The relic density of the DM is given by
\begin{align}
&\Omega h^2
\approx 
\frac{1.07 \times 10^9}{\sqrt{g_*(x_f)} M_{Pl} J(x_f)[{\rm GeV}]},
\label{eq:relic-deff}
\end{align}
where $g^*(x_f \approx 25) \approx 100$, $M_{Pl}\approx 1.22 \times 10^{19}$,
and $J(x_f) (\equiv \int_{x_f}^\infty dx \frac{\langle \sigma v_{\rm rel}\rangle}{x^2})$ is given by
\begin{align}
J(x_f) &= \int_{x_f}^\infty dx\left[ \frac{\int_{4m_\chi^2}^\infty ds\sqrt{s-4 m_\chi^2} (\sigma v_{\rm rel}) K_1\left(\frac{\sqrt{s}}{m_\chi} x\right)}{16  m_\chi^5 x \left[ K_2(x) \right]^2}\right], \nn \\
(\sigma v_{\rm rel}) &= 
\frac{g'^4 q_x^2  s (s-m_\chi^2)}{3\pi \left[ (s-m_{Z'}^2)^2 + m_{Z'}^2 \Gamma_{Z'}^2 \right]},\nn \\
\Gamma_{Z'} &\approx \frac{g'^2 m_{Z'}}{4\pi} + \frac{|g' q_x|^2}{16\pi m_{Z'}^2} (m_{Z'}^2- 4 m_\chi^2)^{3/2}.
\label{eq:relic-deff}
\end{align}
With $s$ is a Mandelstam variable, and $K_{1,2}$ are the modified Bessel functions of the second kind 
of order 1 and 2, respectively. We expect $Z'$ decays into $\mu \bar\mu$, $\tau \bar \tau$,
and $\chi \chi^*$ pairs.~\footnote{Without decay width of $Z'$, the cross section at around the pole of $2m_\chi=m_{Z'}$ is too large and the relic density would be underestimated.} In our numerical analysis, we use the current experimental 
range for the relic density at 3$\sigma$ confidential level~\cite{Ade:2013zuv}:
$0.11 \lesssim \Omega h^2 \lesssim 0.13$.

\noindent \underline{\it Direct detection of DM}:
The dominant elastic scattering cross section arises from the $Q'$ exchange process
in Fig.~\ref{fig:dd}, and its effective Lagrangian of the component level is given by
\begin{align}
\mathscr{L} \simeq - i \sum_{i=1}^{3}  \sum_{a=1}^3 \frac{f^\dag_{ia} f_{a i}}{2 M^2_a}
[\bar q_i \gamma^\mu P_L q_i ][\chi^* \overleftrightarrow\partial_\mu\chi],
\end{align}
where we use the following assumptions of the four transferred momentum $q^2=(p_1-p_2)^2 << M^2_a$ and
the nucleus of a target almost stops (at rest frame);
$\chi^*(p_1) (p_1-p_2)_\mu \chi(k_1)= \frac12 \chi^*(p_1) \left[(p_1-p_2) + (k_1-k_2)\right]_\mu \chi(k_1) \approx \frac12 \chi^*(p_1) (p_1+ k_1)_\mu \chi(k_1)=\frac{i}{2}\chi^* \overleftrightarrow\partial_\mu\chi$,
where the right and left sides correspond to the operators in momentum space and spacetime, respectively.
We then straightforwardly define the DM-nucleon elastic scattering operator as follows:
\begin{align}
\mathscr{L}_N &\simeq  - i  \sum_{i=u}^{d,s}  \sum_{a=1}^3\frac{f^\dag_{ia} f_{a i}}{4 M^2_a} [ F^{q_i/N}_1\bar N \gamma^\mu N
  - F^{q_i/N}_A\bar N \gamma^\mu \gamma_5 N ] \nn \\
&\times (\chi^* \overleftrightarrow\partial_\mu\chi ),
\end{align} 
where we assume that this process is an elastic scattering,
then the four transferred momentum is expressed by $q^2 \approx 0$,
where $p_1$ and $p_2$ are the four momentum of the $\chi^*$ and $q_i or N$, respectively, 
and $F^{q_i/N}_{1,A}$ are the form factors, which is taken from Ref.~\cite{Bishara:2017pfq}.
The squared matrix element is given by
\begin{align}
  \label{eq:infer}
|{\cal M}|^2 &=  \frac{1}{16} 
|\sum_{i=1}^{3}  \sum_{a=1}^3 \frac{f^\dag_{ia} f_{a i} }{M^2_a}
( F_1^{q_i/N} \langle N(k_2) | \bar N \gamma^\mu N | N(p_2) \rangle \nn \\
&-
F_A^{q_i/N} \langle N(k_2) | \bar N \gamma^\mu\gamma_5 N | N(p_2) \rangle ) |^2 \nn \\
&\times |\langle \chi(p_1) | \chi^*\overleftrightarrow\partial_\mu\chi | \chi (k_1) \rangle |^2\nn\\
&\simeq
2 m_N^2 m_\chi^2 \sum_{i=1}^{3}  \sum_{a=1}^3
\left(
\left|\frac{f^\dag_{ia} f_{a i} F_1^{q_i/N}}{M^2_a}\right|^2 + \left|\frac{f^\dag_{ia} f_{a i} F_A^{q_i/N}}{M^2_a}\right|^2
\right),
\end{align}
where the first and second terms in Eq.~(\ref{eq:infer}) do not have an interference term. 

Finally, the complete form of the DM-nucleon elastic scattering cross section is expressed by
\begin{align}
\sigma &\approx \left(\frac{m_\chi m_N}{m_N+m_\chi}\right)^2  \frac{|{\cal M}|^2}{32\pi m_\chi^2m_N^2}\nn\\
&=
\frac1{16\pi }
\left(\frac{m_\chi m_N}{m_N+m_\chi}\right)^2 \nn \\
&\times \sum_{i=1}^{3}  \sum_{a=1}^3
\left(
\left|\frac{f^\dag_{ia} f_{a i} F_1^{q_i/N}}{M^2_a}\right|^2 + \left|\frac{f^\dag_{ia} f_{a i} F_A^{q_i/N}}{M^2_a}\right|^2
\right),
\label{eqq:dd}
\end{align}
where $\sum F_1 \approx 3$ corresponds to the effective operator
$(\bar N \gamma^\mu N )(\chi^* \overleftrightarrow{\partial_\mu}\chi)$, 
$\sum F_A\approx 0.49$ corresponds to the effective operator
$(\bar N \gamma^\mu\gamma_5 N)(\chi^*\overleftrightarrow{\partial_\mu}\chi)$,
and $m_N \approx 0.939$ GeV.
The current experimental upper bounds for the cross section of
the spin independent DM-nucleon elastic scattering are
respectively $\sigma_{\rm exp}\lesssim 2.2 \times 10^{-46}$ cm$^2$ at $m_\chi\approx50$ GeV
for the LUX data~\cite{Akerib:2016vxi}, and $\sigma_{\rm exp} \lesssim 4.1 \times 10^{-47}$ cm$^2$
at $m_\chi \approx 30$ GeV for the XENON data~\cite{Aprile:2018dbl}.
In our numerical analysis, we conservatively restrict
the LUX/XENON1T bounds for the whole range of the DM mass.
%
\begin{figure}[t]
\centering
\includegraphics[width=65mm]{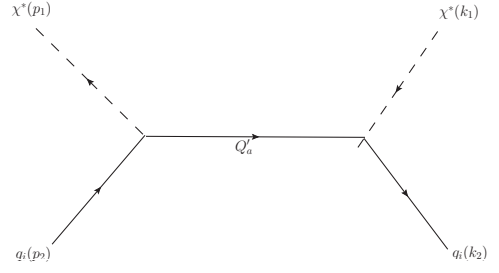}
\caption{A Feynman diagram contributing to the spin independent
  DM-nucleon elastic scattering.} 
\label{fig:dd}
\end{figure}

\section{Numerical analysis} \label{sec:numerical}
In this section, results for our numerical analysis are presented.
In our analysis, we fix a parameter $|q_x| =1 $ for simplicity.
The ranges of the input parameters are set as follows: 
\begin{align}
& 
g'  \in[0.1,1],\ f  \in[0.001,\sqrt{4\pi}],\  m_{Z'}\in [10, 300]\ [{\rm GeV}],\ \nn \\
& m_{\chi} \in [10,150]\ [{\rm GeV}], \ M_{a}\in [{ 1000},3000]\ [{\rm GeV}],
\end{align}
where the lowest DM mass 10 GeV is an assumption to satisfy a
condition $(m_\tau/m_\chi)^2<<1$ in the cross section of relic density.
}
In this calculation, we assume $M_1 < M_2 < M_3$, and $m_{Z'} > m_\chi$.
We then search the allowed region using the range of the input parameters listed above
to satisfy all constraints, namely, {\it $M-\overline M$ mixing,
the measure of the relic density of the DM, the spin independent DM-nucleon scattering
cross section via $Z'$ boson exchange,
and the constraint of the LHC} as well as to explain the anomaly of
$b \to s \bar\mu \mu$ decay.

\begin{figure}[t]
\begin{center}
\includegraphics[width=80mm]{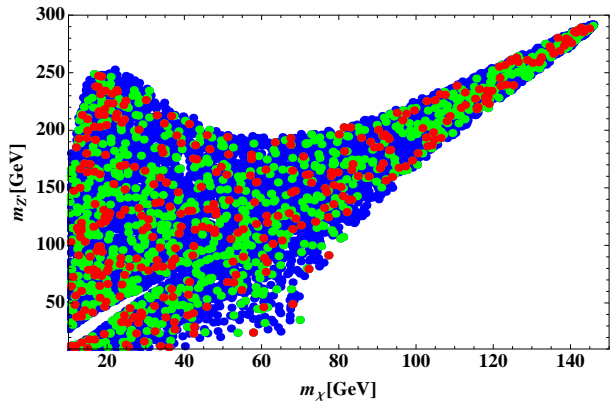}
\caption{The allowed region of $\Delta C_9$ on the plane of $m_\chi$ and $m_{Z'}$
  satisfy the relic density 0.11 $\sim$ 0.13 and the neutral meson mixings, $b \to s \gamma$.
  The blue, red, and green dots are respectively the region of 
$\Delta C_9 \lesssim -5$, $-1.29 \lesssim \Delta C_9 \lesssim -0.87$
at 1$\sigma$, and $-1.67 \lesssim \Delta C_9 \lesssim -0.39$ at 3$\sigma$. 
} 
\label{fig:DM-x-Zp}
\end{center}\end{figure}

Fig.~\ref{fig:DM-x-Zp} shows the allowed region on the plane of $m_\chi$ and $m_{Z'}$,
where the blue, red, and green dots represent respectively the region corresponding to
no constraint on $\Delta C_9$, 1 $\sigma$ range $-1.29 \lesssim \Delta C_9 \lesssim -0.87$,
and 3$\sigma$ range $-1.67 \lesssim \Delta C_9 \lesssim -0.39$.
{\it Note that the lowest point of the maximum absolute value of $f$ is of the order 0.1 for $1(3)\sigma$.}
The correlation between $m_\chi$ and $m_{Z'}$ in Fig.~\ref{fig:DM-x-Zp} comes from
the closed resonant point of the relic density of the DM, when $m_\chi$ is heavier.
In the lighter region of $m_\chi$, the allowed region becomes
wider due to the larger cross section. At around the resonant
region of $10 \ {\rm GeV}\lesssim m_\chi\lesssim 40$ GeV, there are no allowed region
since the corresponding cross section is too large to satisfy the relic density.
This clearly indicates that the mass ranges of the DM and $Z'$ are
respectively 10 GeV $\le m_\chi \lesssim$ 146 GeV and 10 GeV $\le m_{Z'} \lesssim$ 295 GeV,
where the specific ranges mainly originates from
the constraint of the relic density, although the lowest bound of DM
mass 10 GeV comes from the lowest input parameter of the DM mass.
The anomaly of the decay of $b \to s \bar\mu \mu$ is well explained
for the whole allowed region on the plane of $m_\chi$ and $m_{Z'}$.

\begin{figure*}[t]
\begin{center}
\includegraphics[width=80mm]{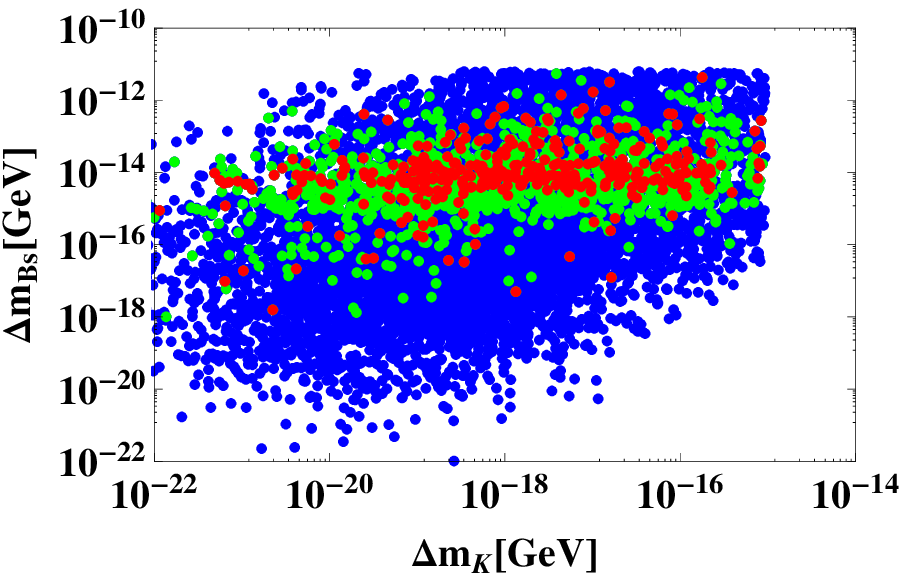}\quad
\includegraphics[width=80mm]{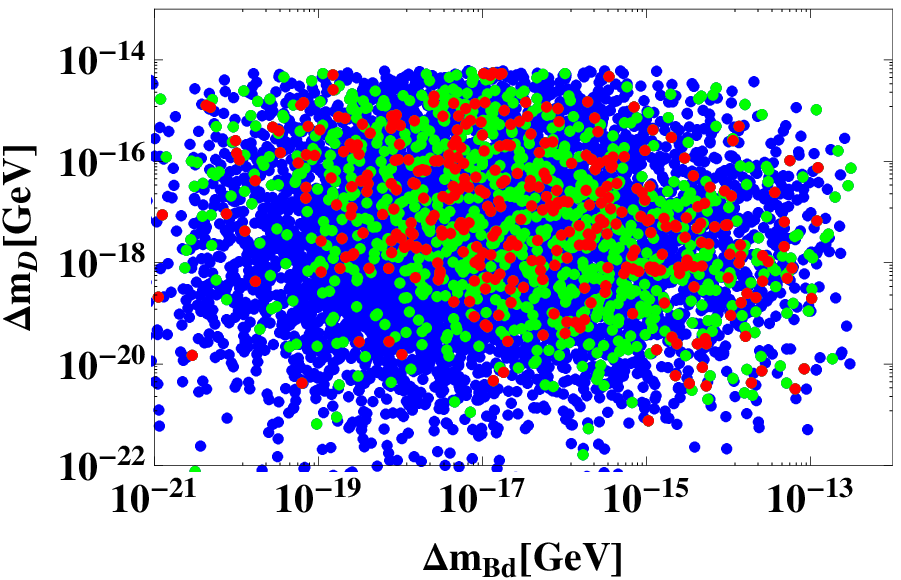}
\caption{The allowed region on the $\Delta m_{K}-\Delta m_{B_s}$ plane (the left panel)
and $\Delta m_{B_d}-\Delta m_{B_D}$ plane (the right panel), where
the color representations are similar as in Fig.~\ref{fig:DM-x-Zp}. } 
\label{fig:m-m}
\end{center}
\end{figure*}

In Fig.~\ref{fig:m-m}, we clearly show that the allowed regions
on the planes of $\Delta m_{K}-\Delta m_{B_s}$ (the left panel) and
$\Delta m_{B_d}-\Delta m_{D}$ (the right panel),
where the color representation is similar as in Fig.~\ref{fig:DM-x-Zp}.
Fig.~\ref{fig:m-m} indicates the allowed region for explaining the anomaly of
$b \to s \bar\mu \mu$, which tends to lie in the range of the experimental bounds
in our parameter space.
 %
A branching ratio of BR (${b \to s \gamma}$) is restricted to be {4.02 $\times$ 10$^{-4}$},
but the typical value is at most of the order $10^{-10}$.
Therefore, the present model clearly satisfies this constraint. Additionally,
we found that the LHC constraint tends to be weaker
than the experimental bounds for the neutral meson mixings in our parameter space.

\begin{figure}[t]
\begin{center}
\includegraphics[width=80mm]{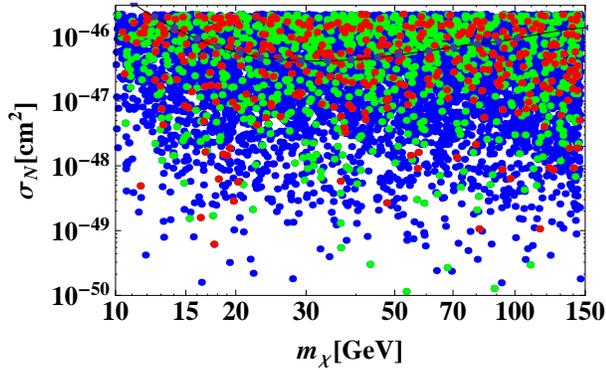}
\caption{The nucleon-DM elastic scattering generated
by the parameter space scanned. The current upper bound by XENON-1T
is represented by the solid line.} 
\label{fig:DM-dd}
\end{center}\end{figure}

Fig.~\ref{fig:DM-dd} shows the cross section of
the nucleon-DM elastic scattering obtained from our parameter space scanned.
This indicates that the parameter points excluded by
the present XENON1T data~\cite{Aprile:2018dbl}. 
This means more parameter points can be explored in the future experiments. 

\section{Summary and Conclusions} \label{sec:summary}
We have explored the possibility of explaining the experimental anomalies
in the semileptonic decay of the $B$-meson, $B \to K^{*} \mu \bar\mu$,
in the framework of the gauged $U(1)_{\mu-\tau}$ symmetry.
With our present model, which is built in a more complete manner than previous model,
we have performed a more detailed analysis by searching the allowed region
from several present experimental constraints.

Apart from the muon anomalous magnetic moment and neutrino sector,
we have formulated a model starting with a valid Lagrangian
by considering the Higgs potential with the inert conditions,
the Wilson coefficient for the decay of $B \to K^{*} \mu \bar\mu$,
the collider physics, the neutral meson mixings, the bound on direct detection,
and the relic density of a bosonic dark matter candidate.
We have searched the parameter space, which explain the size of
the anomaly of $B \to K^{*} \mu \bar \mu$ decay, satisfying all constraints.
We found that the allowed region on the plane of the DM and $Z'$ masses is narrower
compared to the previous analysis in the heavier DM mass. This is expected due to
the decay width of $Z'$ is rather large.
On the other hand, in the lighter region of $m_\chi$, the allowed region becomes to be still wider because the cross section is larger.
Moreover, there are no allowed region at around the resonant region of $10 \ {\rm GeV}\lesssim m_\chi\lesssim 40$ GeV,
since the corresponding cross section is too large to satisfy the relic density.
The meson mixing of $\Delta B_s$ can be well tested in the future experiments,
since the structure of the Yukawa couplings $f$ is the same as the one of
$\Delta C_9$ of $B \to K^{*} \mu \bar\mu$.
The absolute parameter of $f$ can naturally be estimated to explain $\Delta C_9$,
and its minimum absolute is at most of the order $\sim$ 0.1.


\section*{Acknowledgments}
H.O. was supported by the Ministry of Science, ICT and Future Planning,
Gyeongsangbuk-do and Pohang City through the Junior Research Group (JRG) of APCTP.
The work of Y.O. was supported from European Regional Development
Fund-Project Engineering Applications of Microworld Physics (No.CZ.02.1.01/0.0/0.0/16-019/0000766).
The work of P.T.P.H. was supported by the Ministry of Science, ICT and Future Planning,
Gyeongsangbuk-do and Pohang City through
the Asia Pacific Economic Cooperation-Young Scientist Training (APEC-YST) of APCTP.
H.O. is sincerely grateful for the Korea Institute for Advanced Study (KIAS) and all the members.

\end{document}